\documentclass[a4paper]{jpconf}

\usepackage{graphicx}
\usepackage{bm}
\usepackage[latin1]{inputenc}
\usepackage{color}
\usepackage{amssymb}
\usepackage{times}
\usepackage{amsmath}
\usepackage{colordvi}
\usepackage{color}
\usepackage{epsfig}
\usepackage{upgreek}

\usepackage{cite}

\begin{document}

\title{Status of the GEO\,600 squeezed-light laser}

\author{Alexander Khalaidovski, Henning Vahlbruch, Nico Lastzka, Christian Gr\"af, Harald L\"uck, Karsten Danzmann, Hartmut Grote and Roman Schnabel}
\address{Max-Planck-Institut f\"ur Gravitationsphysik (Albert-Einstein-Institut) and\\
\mbox{Institut f\"ur Gravitationsphysik, Leibniz Universit\"at Hannover}\\
Callinstr.~38, 30167 Hannover, Germany}
\ead{Roman.Schnabel@aei.mpg.de}

\date{\today}

\begin{abstract}
In the course of the high-frequency upgrade of GEO\,600, its optical configuration was extended by a squeezed-light laser~\cite{VKLGDS10}. Recently, a non-classically enhanced measurement sensitivity of GEO\,600 was reported~\cite{SqzNatPhys}. In this paper, a characterization of the squeezed-light laser is presented. Thereupon, the status of the integration into GEO 600 is reviewed, focussing on the sources of optical loss limiting the shot noise reduction by squeezing at the moment. Finally, the possibilities for a future loss reduction are discussed.
\end{abstract}

\section{Introduction}
The German-British project GEO\,600~\cite{G10} is part of an international network of kilometre-scale Michelson-type laser interferometers~\cite{A09, A08} searching for gravitational wave (GW) signals at audio frequencies in a band of 10\,Hz\,--\,10\,kHz. While at frequencies below a few hundred Hertz the measurement sensitivity is, at the moment, still limited by seismic, thermal and technical noises, at higher frequencies photon shot noise constitutes the limiting factor (for an extended review of quantum effects in GW interferometers, see~\cite{SMML10}). Currently, GEO\,600 undergoes an upgrade program to GEO\,-\,HF, aiming at a measurement sensitivity improvement by about an order of magnitude at high frequencies~\cite{Wetal06, Letal10}.

A gravitational wave signal is `translated' by the laser interferometer into a variation of the optical power at the signal output port. A limit for the possible measurement accuracy is set by the Poissonian distribution of the photon number arriving within a time interval at the interferometric photo detector. Laser GW interferometers are operated close to their \emph{dark fringe}, which means that due to destructive interference almost no light leaves the signal port and all optical power is reflected back to the laser. However, zero-point (vacuum) fluctuations of the electro-magnetic field are coupled into the interferometer's signal port and generate a signal that is not distinguishable from a GW signal. A possibility to reduce this \emph{shot noise}, as was first proposed by Caves in~\cite{C81}, is the use of \emph{squeezed light}~\cite{Y76, GK04}. Such a light field has a non-classical noise distribution with a reduced uncertainty in one of the field quadratures, when compared to a vacuum state. Injected from the signal port, the squeezed state replaces the vacuum state, thereby directly reducing the interferometer's quantum noise.

Since the first observation of squeezed light in the 1980s~\cite{SHYMV85, WXK87, PXKH88}, the degree of squeezing has been constantly improved~\cite{Vetal08, MVLDS10, Mehmet12}, recently reaching a value of almost 13\,dB~\cite{Eberle10}. In the GW detection band, the generation of squeezing remained an unsolved problem for a long time. It was not until 2004 that squeezing at audio frequencies down to 280\,Hz could be generated for the first time~\cite{MGBWGML04}, soon followed by the creation of a coherent control scheme~\cite{Vahlb06, Cetal07} that allowed the observation of frequency-independent squeezing over the entire Earth-bound gravitational wave detection band from 10\,Hz to 10\,kHz~\cite{Vahlb07}. The implementation of squeezed light was tested in a table-top Michelson-type interferometer~\cite{Vahlb07, MSCBL02} as well as in a suspended GW prototype detector \cite{GMMetal08}. Finally, a squeezed-light laser for GEO\,600 was constructed~\cite{VKLGDS10}, first results of a squeezing-improved measurement sensitivity were recently published in \cite{SqzNatPhys}. In this paper, a characterization of the squeezed-light laser is presented along with the current status of the implementation in GEO\,600.

\section{Optical layout}
\begin{figure}[t]
\begin{center}
		\includegraphics[width=0.9\linewidth]{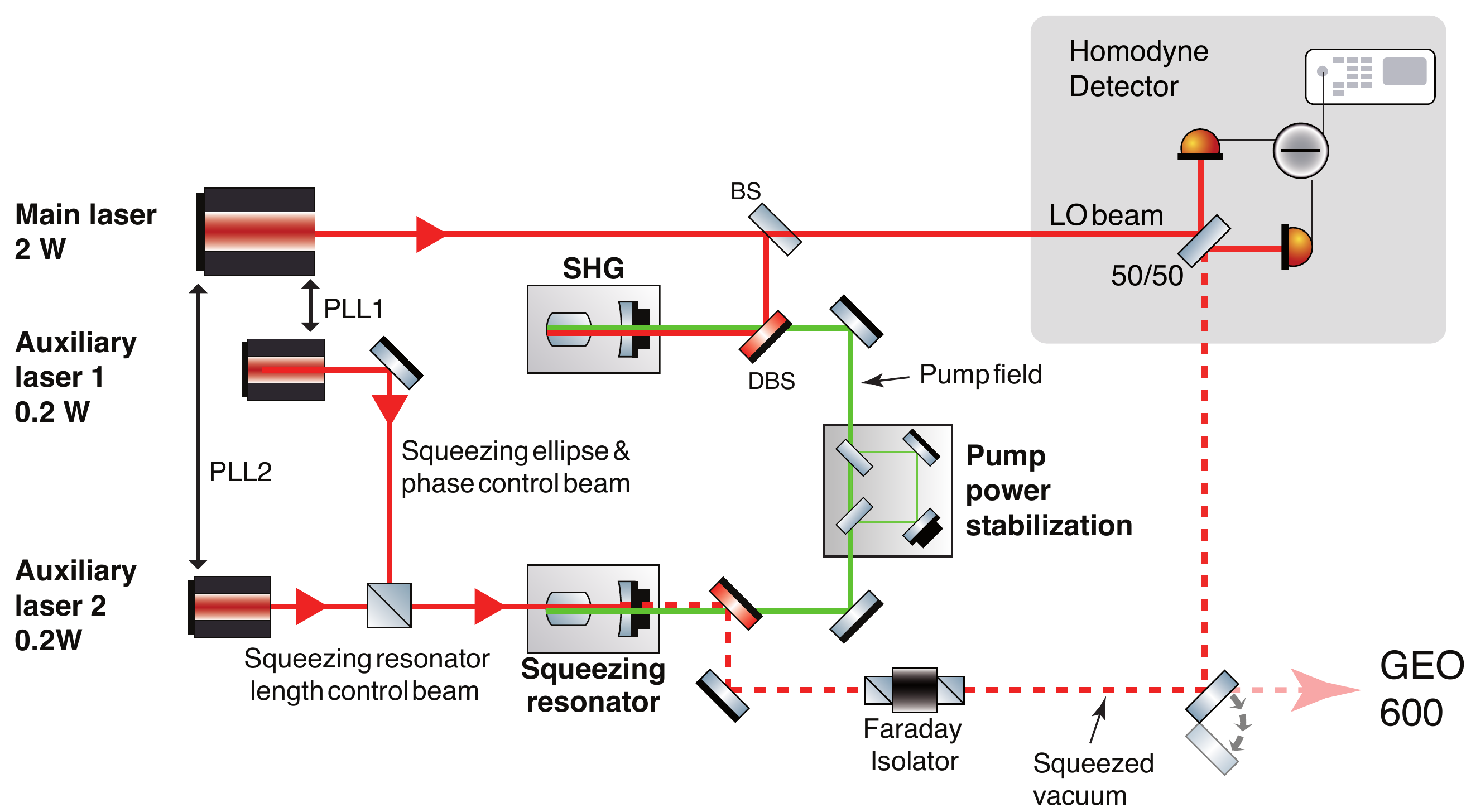}
\end{center}
		\caption{Simplified sketch of the optical layout of the squeezed-light laser. Auxiliary laser 1 is used to stabilize the length of the squeezing resonator and auxiliary laser 2 allows controlling the phase of the squeezed field. The two auxiliary lasers are phase locked (PLL) to the main 2\,W laser. For characterization, a balanced homodyne detector is available.}
\label{figure1}
\end{figure}

Figure~\ref{figure1} shows a schematic of the optical layout of the squeezed-light laser. To maintain clarity, only the core components are shown. Altogether, the experiment consists of approximately 130 components which are situated on a 135\,cm $\times$ 113\,cm optical breadboard. A detailled discussion of the single subsystems and of the employed control scheme is provided in Reference~\cite{VKLGDS10}.

The experiment is driven by a monolithic non-planar Nd:YAG ring laser (NPRO) of 2\,W single-mode output power at 1064\,nm. One part of this beam is frequency up-converted in a second-harmonic generator (SHG), which uses 7\% MgO doped LiNbO$_3$ as nonlinear $\chi^{(2)}$ medium. This 532\,nm \emph{pump beam} is filtered by a mode-cleaning travelling-wave resonator (not shown in Fig.~\ref{figure1}). Thereby, high-frequency phase noise that has been shown to diminish the maximal degree of squeezing achievable~\cite{TYYF07, FHDFS06}, is attenuated. Subsequently, the pump beam is injected into the squeezed-light source that consists of a periodically poled potassium titanyl phosphate (PPKTP) crystal placed in a standing-wave hemilithic cavity.

The usage in a gravitational wave detector requires a high long-term stability of the squeezing degree and consequently a high stability of the pump beam power. The dependence of the degree of squeezing on the pump power value is discussed in detail in~\cite{KVLGDGS11}. A Mach-Zehnder interferometer is employed to stabilize the pump beam intensity. The scheme ensures the squeezing value to be stable on long-term timescales, first characterization results are presented in~\cite{KVLGDGS11}.

The generation of squeezed states at audio frequencies from 10\,Hz to 10\,kHz requires the implementation of a control scheme that avoids the introduction of technical laser noise to the squeezed vacuum state. The basic idea is to remove all noisy fields that could by means of interference introduce noise to the squeezed field at the frequencies of interest~\cite{ MGBWGML04, BSTBL02, SVFCGBBLD04}. The \emph{coherent control scheme} \cite{Vahlb06, Cetal07} employed uses two auxiliary optical frequencies that are provided by two additional NPRO laser sources. The frequency of the auxiliary units is locked to the main laser that is in turn phase-locked to the GEO\,600 master laser when operating the squeezed-light laser at the detector site. 

For characterization of the squeezed-light laser, an on-board diagnostic homodyne detector is available. A small fraction of the main 1064\,nm beam, that is also filtered by a ring mode-cleaner, is used as a local oscillator (LO) beam. 

\section{Characterization of the squeezed-light laser}
\begin{figure}[t]
\begin{center}
		\includegraphics[width=\linewidth]{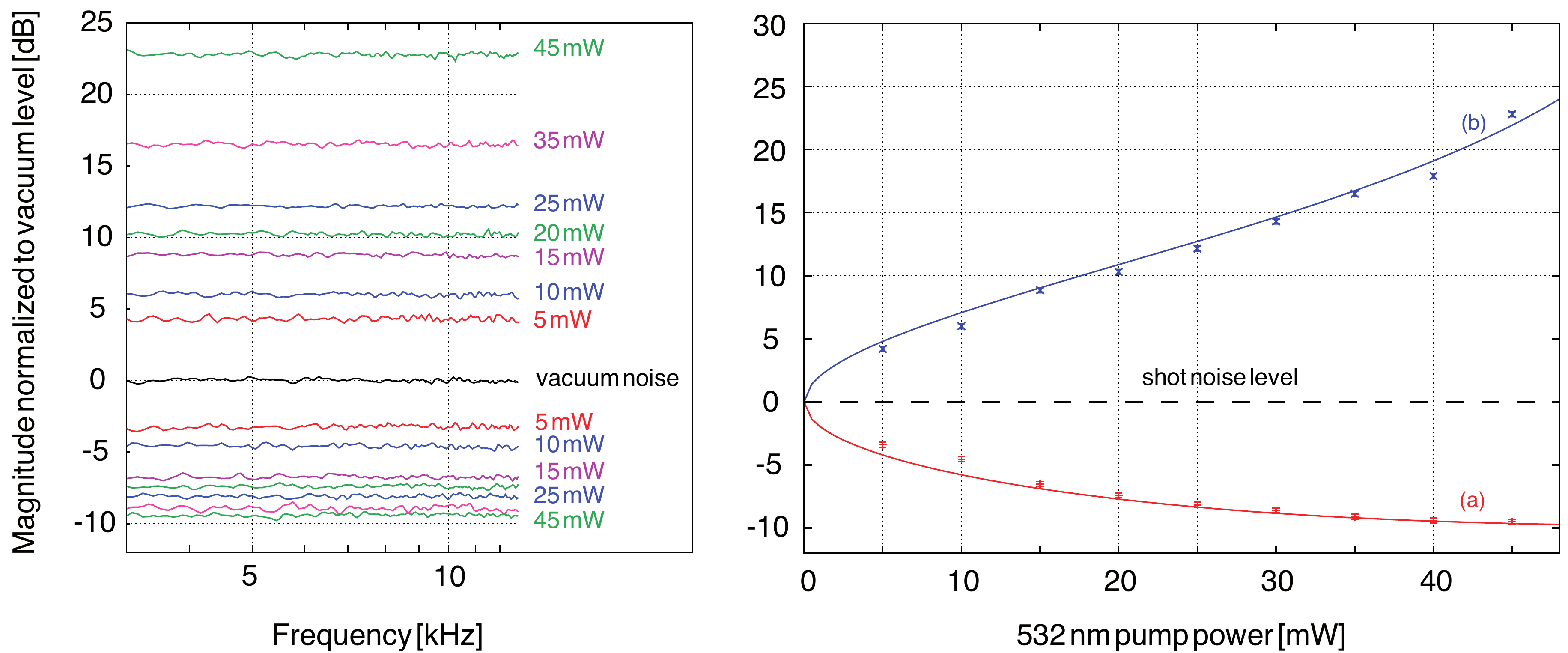}
\end{center}
		\caption{Left: Squeezed and anti-squeezed variances measured at different pump powers.  A threshold power of 61\,mW was estimated from the measurements. For clarity, only a small frequency range of 3\,kHz\,--\,12.8\,kHz is shown. Right: the blue and red points show the squeezing and anti-squeezing measured at 5\,kHz, as shown in the left graph. Traces (a) and (b) show the theoretical prediction of the expected anti-squeezing and squeezing levels, respectively, taking into account an overall loss of 10.5\,\%.}
\label{figure2}
\end{figure} 

The left part of Figure~\ref{figure2} shows squeezing and anti-squeezing values measured for different 532\,nm pump powers. The values are normalized to the noise level of a vacuum state that can be measured when the signal port of the diagnostic homodyne detector is blocked (black trace). It is worth noting that very low 532\,nm pump powers are sufficient to produce a high degree of squeezing. At a power of 5\,mW (corresponding to about 15\,mW intra-cavity power), the quantum noise was reduced by more than 3\,dB, equivalent to a factor of two in noise power reduction. The degree of squeezing reached a maximal value of 9.5\,dB at a pump power of 45\,mW. This value is more than an order of magnitude below the pump power levels necessary to reach correspondent squeezing values with similar cavity designs using MgO:LiNbO$_3$ as nonlinear medium~\cite{Vetal08, MVLDS10}. To the best of our knowledge, the observed non-classical noise suppression of 9.5\,dB sets a new benchmark for the degree of squeezing up to now reported at audio frequencies.

In the right part of Figure~\ref{figure2}, the measured values (taken at a frequency of 5\,kHz) are compared with the theoretical expectations. The squeezed and anti-squeezed noise powers in decibel \cite{TYYF07}
\begin{equation}
V_\text{s/a} = 10\cdot\text{log}_{10}(R_\text{s/a})
\label{eq.ressqz1}
\end{equation}
with the variances
\begin{equation}
R_\text{s/a} = 1\mp \eta_\text{opt} \frac{4\sqrt{P/P_\text{th}}}{(1 \pm \sqrt{P/P_\text{th}})^2+4\kappa^2}
\label{eq.mz1}
\end{equation}
depend on the pump power $P$ as well as on optical loss. In this expression, \textit{a} and \textit{s} and the upper and lower signs denote the anti-squeezing and the squeezing, respectively, $\eta$ describes the total detection efficiency and $P_\text{th}$ is the OPO threshold power. The normalized frequency $\kappa$ that is a function of cavity parameters is not relevant for the presented discussion, since frequencies much smaller than the cavity linewidth are considered.\\[0.8ex]
\noindent The different loss sources during the generation and detection proscesses are:
\begin{itemize}
\item The escape efficiency $\rho\,=\,T/(T+L)$ of the squeezed-light resonator. Here, $T\,=0.08$ is the power transmittance of the output coupling mirror and $L$ is the intra-cavity loss. The total loss per round-trip, due to the non-perfect AR-coating of the crystal facet as well as to the residual power absorption of the nonlinear crystal, can be estimated to $L\,\approx 0.04$, thus resulting in $\rho\approx 0.95$\,.
\item The total propagation loss suffered by the squeezed field on the way from the squeezing resonator to the homodyne detector. The critical components in this path are two polarizing beam splitters as well as a Faraday rotator, while the contribution of the super-polished lenses is negligible. In an independent measurement, the propagation efficiency was estimated to be $\xi \approx 0.98$\,.
\item The fringe visibility $\mathcal{V}$ at the homodyne detector, contributing quadratically, was measured to range between $0.985$ and $0.9875$.
\item The quantum efficiency of the homodyne photo diodes. To minimize this loss, custom-made high-efficiency diodes are used. The diameter of the active area is $500\,\upmu$m. Equivalent diodes have already been used in high-level squeezing expe\-ri\-ments, as e.\,g.~reported in \cite{Vetal08, MVLDS10}, and have been estimated to have a quantum efficiency of $\eta_\text{hmd}>0.99$\,.
\end{itemize}
A consideration of all possible sources results in a total detection efficiency of \mbox{$\eta_\text{tot}\approx 0.895$}. This value corresponds very well to the measured squeezing and anti-squeezing levels, as shown in Figure~\ref{figure2}. From the measurements, an OPO threshold power of $P_\text{th}\approx 61$\,mW was estimated.\\[0.8ex]
It is important to note that the homodyne detector is merely a diagnostic device which is by-passed when the squeezed-light laser is used in GEO\,600. Hence, the homodyne visibility loss is not relevant for a later operation and has to be corrected for. Thus, a total loss of merely 7\,\% can be deduced. 

The initial requirement formulated for the squeezed-light laser was a squeezing value of 10\,dB. When operated at a pump power of 35\,mW, more than 10\,dB of squeezing are available for the injection into the dark GEO\,600 port. The advantage of this operation point is that with a corresponding anti-squeezing value of about 17\,dB, the generated state is still relatively pure. If the pump power is increased further, the gain in squeezing is marginal while the anti-squeezing increases rapidly, leading to an increased sensitivity to phase noise. The highest squeezing value available is 11.3\,dB (with about 23\,dB of anti-squeezing) when using a pump power of 45\,mW. The high squeezing factor was achieved over almost the entire frequency band of Earth-bound instruments extending from 10\,Hz up to 10\,kHz~\cite{VKLGDS10}. The long-term stability of the device was discussed in~\cite{KVLGDGS11}. Altogether, this means that the requirements initially formulated for the GEO\,600 squeezed light laser have been fully achieved.

\section{The squeezed-light laser in GEO\,600}
\label{sqzingeo}
\begin{figure}[t]
\begin{center}
		\includegraphics[width=\linewidth]{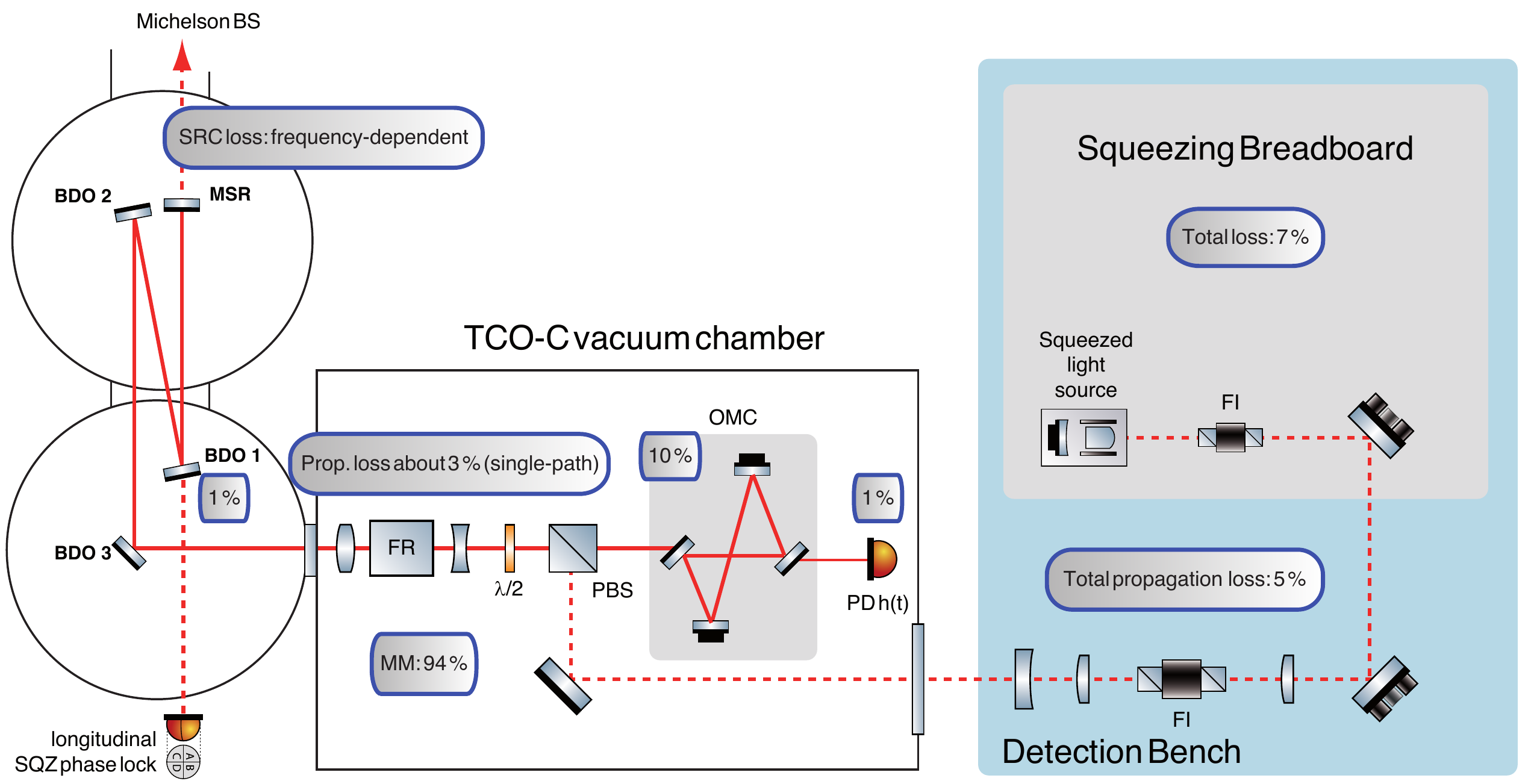}
\end{center}
		\caption{Sketch of the optical layout of the GEO\,600 detection stage with squeezing input. Auxiliary optics are omitted for clarity reasons. The text boxes show estimations of the optical loss experienced by the squeezed field.}
\label{figure5}
\end{figure}

A sketch of the squeezing-input stage at the GEO\,600 detector site is shown in Fig.~\ref{figure5}. The squeezing injection scheme is realized through the combination of a Faraday rotator (FR) and a polarizing beam splitter: The squeezed field is reflected at the PBS between the output mode-cleaner (OMC) and the signal-recycling mirror (MSR), experiences a 45 degree polarization rotation passing the FR and another 45 degree rotation after having been reflected at the MSR. The total polarization rotation of 90 degrees allows the squeezed beam to be transmitted by the PBS together with the output beam of the interferometer (carrying the GW signal) and to arrive at the GW signal photo detector after a transmission through the OMC. The squeezed-light laser is situated on the so-called \emph{detection bench}, an optical table located close to the vacuum chamber that contains the squeezing input PBS, the OMC and the GW photo detector (this vacuum chamber is referred to as \emph{TCO-C} in the following).

\begin{figure}[t]
\begin{center}
		\includegraphics[width=0.7\linewidth]{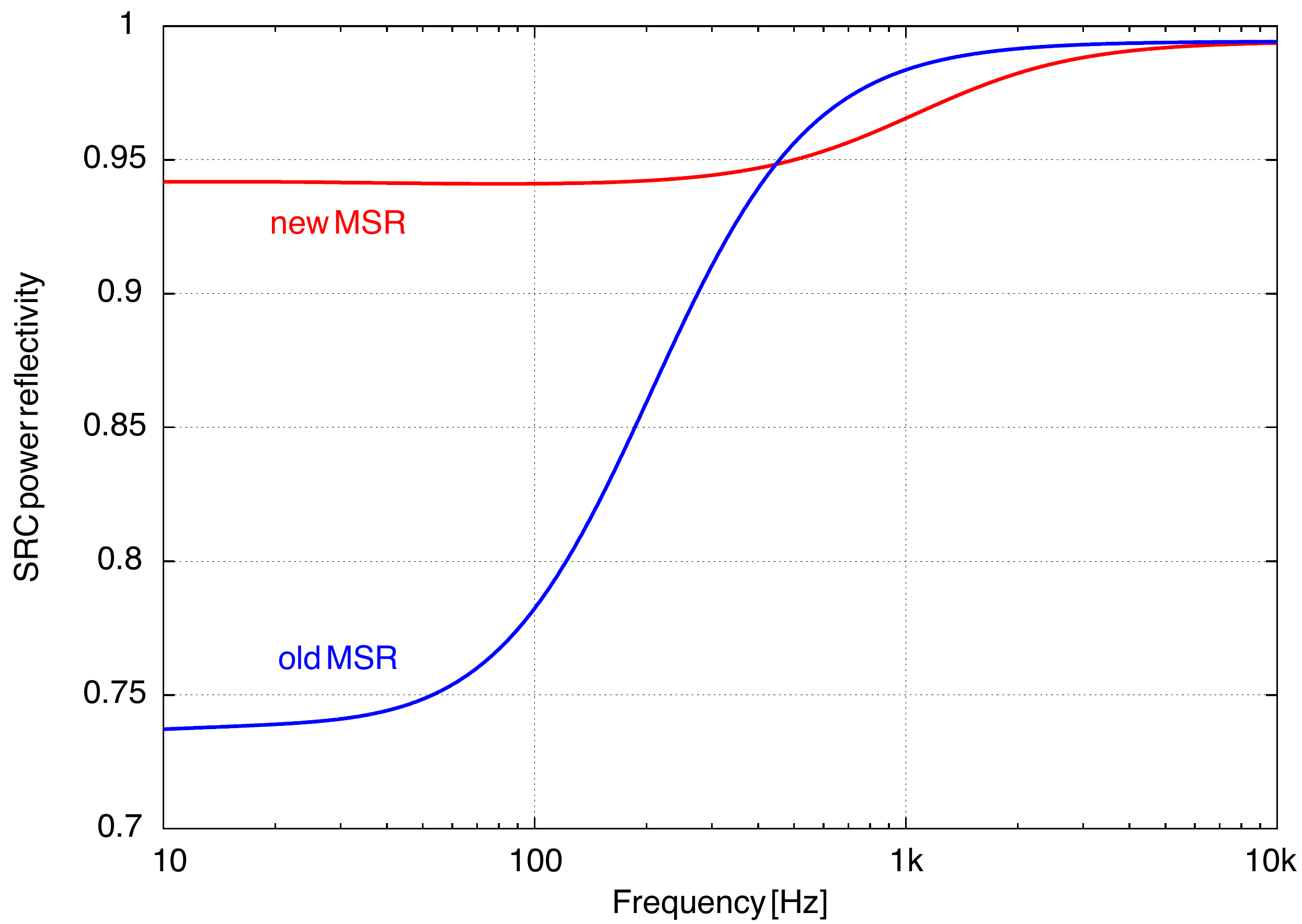}
\end{center}
		\caption{{\sc finesse} simulation of the signal-recycling cavity's frequency-dependent power reflectivity with the old (blue trace) and new (red trace) MSR. The bandwidth of the SRC was increased from 220\,Hz to 1100\,Hz.}
	\label{figure6}
	\end{figure}

At the output of the squeezing breadboard, up to 11.3\,dB of squeezing is available for injection into GEO\,600. However, different loss sources appearing at various stages of the squeezed light injection have to be considered. In the end, they limit the maximal sensitivity improvement achievable with squeezed light. In the following, the optical losses experienced by the squeezed field are summarized, the individual contributions are illustrated in Fig.~\ref{figure5}.
\begin{itemize}
\item The total loss on the squeezing breadboard is, as stated above, about 7\,\%, taking into account an estimated escape efficiency of 95\,\% and a propagation loss of 2\,\%. 
\item During the implementation of squeezing, it was discovered that an additional Faraday isolator was required to protect the interferometer from back-scattering as well as the squeezed-light source from a bright beam coming from the interferometer. The additional Faraday isolator adds 44\,dB of isolation but contributes 4\,\% to the propagation loss. The total propagation loss on the detection bench is therefore about 5\,\%. 
\item The expected single-path propagation loss inside TCO-C, due to imperfect AR coatings and absorption of the PBS, the Faraday rotator and the waveplate can be estimated to be about 3\,\%.
\item 1\,\% is transmitted at one of the suspended steering mirrors (BDO$_1$ in Fig.~\ref{figure5}) for error signal generation (This loss also contributes twice due to the double-path of the squeezed beam).
\item A frequency-dependent loss is introduced by the signal-recycling cavity. While at frequencies outside the SRC linewidth it acts as a mirror, at low frequencies the squeezed field is coupled into the interferometer and experiences the intra-cavity loss. Figure~\ref{figure6} shows the expected power reflectivity of the SRC for the squeezed field. For this, the {\sc finesse} file of GEO\,600 published in~\cite{HildPhD} was used. 

In the course of the GEO\,-\,HF upgrade, the MSR reflectivity was reduced from 98.1\,\% to 90\,\%\,. As far as the squeezed field is concerned, the higher bandwidth leads to a slightly increased loss value at kHz-frequencies. The expected effect is shown in the blue trace of Fig.~\ref{figure6}.
\item The best mode-matching to the signal-recycling cavity and to the output mode-cleaner measured was about 94\,\%.
\item From the design parameters, the optical loss for a single transmission through the OMC was estimated to be below 1\,\%. During characterization, however, a much higher value of about 10\,\% was measured. This loss could not be reduced further by cleaning and probably arises from a combination of imperfect optical surfaces and dielectric coatings deviating from the design specifications. 
\item Before the GEO\,-\,HF upgrade, a Perkin Elmer C30642 photo diode was used. It was substituted by a custom-made high-efficiency photo diode, as already used in the diagnostic homodyne detector. The diameter of the active area is 3\,mm. The quantum efficiency was measured to be 8\,\% higher than for the Perkin Elmer PD and is expected to be similar to the $500\,\upmu$m diodes used in previous squeezing experiments, namely about 99\,\%.
\end{itemize}

\begin{figure}[tb]
\begin{center}
		\includegraphics[width=0.75\linewidth]{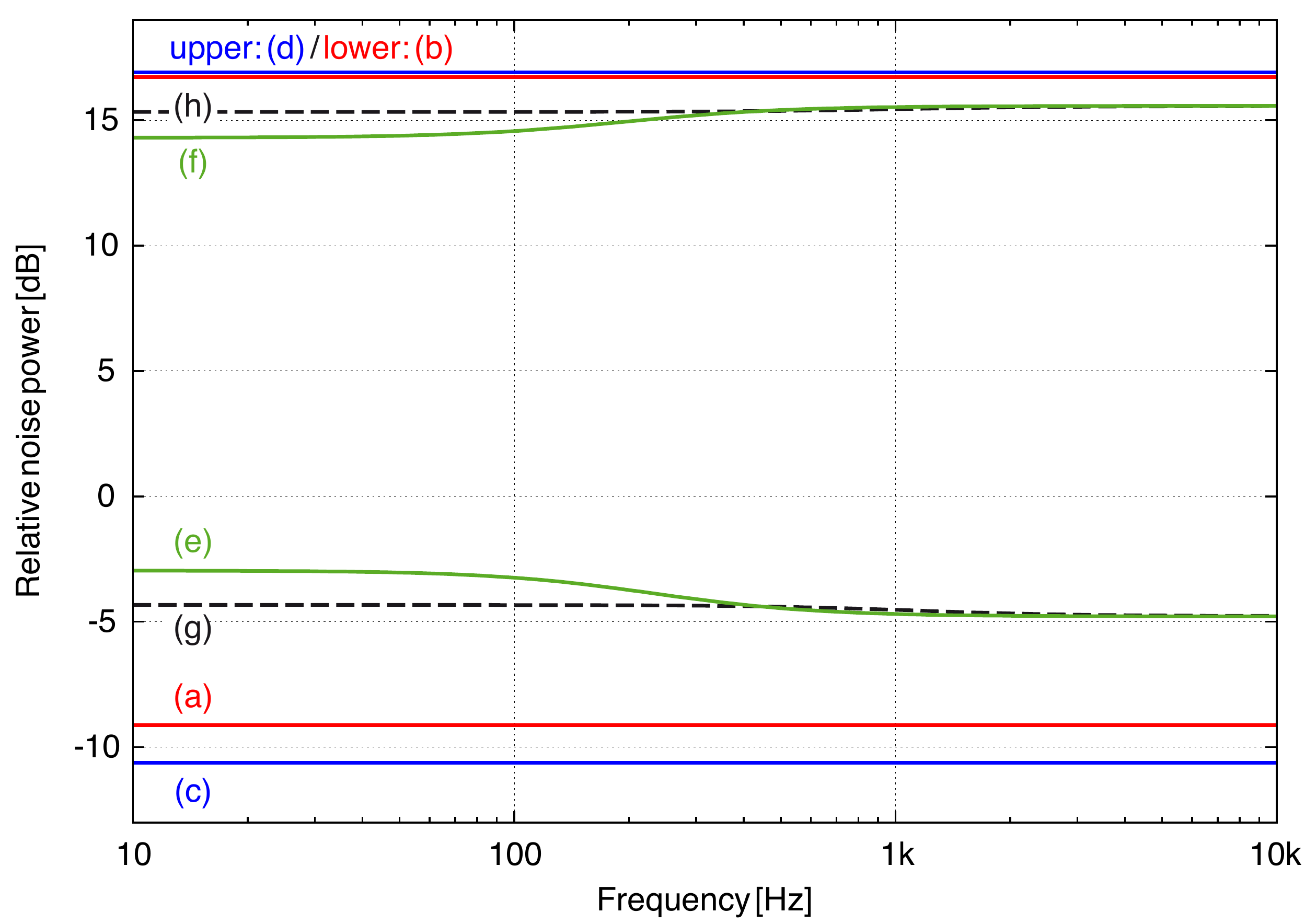}
\end{center}
		\caption{Simulated squeezing and anti-squeezing values prior to and after the injection into GEO\,600. A value of 35\,mW was assumed for the 532\,nm pump power. Traces (a\,--\,b): measuring at the diagnostic homodyne detector. Traces (c\,--\,d): squeezing and anti-squeezing available at the output of the squeezing breadboard, corrected for the homodyne detector losses. Traces (e\,--\,f): frequency-dependent squeezing and anti-squeezing values at the GEO\,600 GW photo detector with the old MSR. Traces (g\,--\,h): same with the new MSR.}
\label{figure7}
\end{figure}

The total loss for the squeezing, being the sum of all contributions mentioned above, is therefore estimated to a value of 32\,\% at kHz-frequencies and up to 50\,\% at frequencies much lower than the SRC linewidth. With 220\,Hz the latter is, however, already in the frequency range where the interferometer sensitivity is no longer limited by quantum noise. Figure~\ref{figure7} shows the expected squeezing and anti-squeezing values for the squeezed-light laser operated with a pump power of 35\,mW. Traces (a\,--\,b) refer to the squeezing and anti-squeezing measured at the diagnostic homodyne detector. Traces (c\,--\,d) correct those values for the homodyne detector loss and therefore show the squeezing or anti-squeezing available for injection at the squeezing breadboard output. Traces (e\,--\,f) finally give an estimation of the amount of squeezing or anti-squeezing present at the GEO\,600 GW photo detector. Traces (g\,--\,h) illustrate the effect of the MSR exchange, leading to an increased SRC linewidth. 

\section{Future work}
The long-term goal to be achieved with squeezed light injection is a desired reduction of the observatory noise by 6\,dB at quantum-noise limited frequencies and a permanent squeezing contribution during the operation of GEO\,-\,HF. After the first successful enhancement of the observatory sensitivity by squeezed light, the achievement of both intents seems to be feasible in the near future. Regarding the squeezing factor, the following loss sources need to be addressed:
\begin{itemize}
\item The optical loss during beam propagation on the detection bench is mainly due to the currently used Faraday isolator with a suboptimal performance in this respect as well as to polarization optics. The use of a high-throughput Faraday isolator identical to the one already used in the squeezed-light laser setup will allow a reduction of the propagation loss. Apart from the Faraday isolator, merely super-polished optics with a high-performance AR-coating (as used on the squeezing breadboard) will be used, thereby reducing loss due to surface scattering and to residual AR-coating reflection.
\item Lenses with a super-polished surface and a low residual reflection at the AR coating (as already used in the squeezed-light laser setup) will be used inside TCO-C.
\item Up to now, the largest single loss contribution arises from the transmission through the output mode-cleaner. In the future, the OMC optics will be exchanged. From the specifications of the single optics, the total transmission loss (mainly due to coating and surface imperfections and to a 100\,ppm residual transmission of one of the mirrors for monitoring purposes) was estimated to be below 1\,\%.
\end{itemize}

An important issue to be addressed is the automatic alignment of the squeezed beam to the interferometer. Without this control loop, the suspended interferometer optics, though locked with respect to each other, are free swinging with respect to the squeezed beam injection path. This introduces a time-dependent, unknown effective loss for the squeezing injection. The generation of auto-alignment error signals is currently under investigation.

\section{Conclusion}
In this paper, we presented a characterization of the GEO\,600 squeezed-light laser. The highest directly observed degree of squeezing is 9.5\,dB, which constitutes the highest degree of squeezing reported so far at audio frequencies. Thus, more than 11\,dB of squeezing is available for the injection into the signal port of GEO\,600. Furthermore, individual sources of optical loss experienced by the squeezed field in the course of generation, propagation in GEO\,600, and the final photo-electric detection at the output port were discussed. The presented possibilities to reduce the total optical loss suggest an even stronger non-classical sensitivity improvement of GEO\,600 by squeezed light to be feasible in the near future.

\ack
This work has been supported by the international Max Planck Research School (IMPRS) and the cluster of excellence QUEST (Centre for Quantum Engineering and Space-Time Research).

\end{document}